\documentclass[a4paper,11pt]{article}

\usepackage{lmodern} 
\usepackage{float}
\usepackage[T1]{fontenc}
\usepackage[utf8]{inputenc}
\usepackage{cite}
\usepackage{mathtools} 
\usepackage{amsfonts}
\usepackage{amsmath} 
\allowdisplaybreaks
\usepackage{dsfont}
\usepackage{physics}
\usepackage[ruled]{algorithm2e}
\usepackage{enumerate}
\usepackage{enumitem}
\usepackage[font=small,labelfont=bf]{caption}
\usepackage{tikz}
\usetikzlibrary{hobby} 
\usepackage{subcaption}
\usepackage{graphicx}
\usepackage[normalem]{ulem} 
\usepackage{array} 
\newcolumntype{C}[1]{>{\centering\let\newline\\\arraybackslash\hspace{0pt}}m{#1}}
\usepackage{color}
\newcolumntype{L}[1]{>{\raggedright\let\newline\\\arraybackslash\hspace{0pt}}m{#1}}

\usepackage{placeins}
\usepackage{changepage} 
\usepackage{multirow}

\usepackage[colorlinks=true,linkcolor=blue,citecolor=citegreen]{hyperref}
\usepackage[affil-it]{authblk}
\usepackage{amsthm}
\usepackage[nameinlink,capitalise]{cleveref}
\newtheorem{theorem}{Theorem}

\crefname{section}{Section}{Sections}
\crefname{subsection}{subsection}{subsections}
\crefname{theorem}{Theorem}{Theorems}
\crefname{corollary}{Corollary}{Corollaries}
\crefname{lemma}{Lemma}{Lemmas}
\crefname{appendix}{Appendix}{Appendices}
\crefname{definition}{Definition}{Definitions}
\crefname{equation}{eq.}{eqs.}
\crefname{algorithm}{Algorithm}{Algorithms}
\definecolor{mygreen}{RGB}{0,150,0}
\definecolor{mygrey}{RGB}{100,100,100}
\definecolor{myred}{RGB}{221,0,0}
\definecolor{tk}{RGB}{246,76,246}
\definecolor{citegreen}{RGB}{0,165,0}

\renewcommand{\ket}[1]{| #1 \rangle}

\newcommand{\set}[1]{\left\{ #1 \right\}}
\newcommand{\1}{\mathds{1}}
\newcommand{\wt}[1]{\widetilde{#1}}
\newcommand{\mc}[1]{\mathcal{#1}}

\newcommand{\Fnorm}[1]{\norm{#1}_\mathrm{F}}
\newcommand{\Flip}{\mathds{F}}

\newcommand{\e}{\mathrm{e}}
\renewcommand{\i}{\mathrm{i}}
\newcommand{\dt}{\mathrm{d}t}
\newcommand{\avg}{\mathrm{avg}}

\DeclareMathOperator{\poly}{poly}
\DeclareUnicodeCharacter{2009}{\,}

\newcommand{\C}{\mathbb{C}}

\begin{document}

\title{Fitting time-dependent Markovian dynamics to noisy quantum channels}

\author[1,2]{Emilio Onorati}
\author[1,3]{Tamara Kohler}
\author[1]{Toby S. Cubitt}
\affil[1]{University College London, Department of Computer Science, UK}
\affil[2]{Technische Universit\"at M\"unchen, Fakult\"at f\"ur Mathematik, DE}
\affil[3]{Instituto de Ciencias Matem\'aticas, Madrid, ES}

\date{}

\maketitle

\begin{abstract}
Understanding how to characterise and mitigate errors is a key challenge in developing reliable quantum architecture for near-term applications.
Recent work~\cite{noMarkpaper21} provides an efficient set of algorithms for analysing unknown noise processes requiring only tomographic snapshots of the quantum operator under consideration, without the need of any a-priori information on the noise model, nor necessitating a particular experimental setup.
The only assumption made is that the observed channel can be approximated by a time-independent Markovian map, which is a typically reasonable framework when considering short time scales.

In this note we lift the time-independent assumption, presenting an extension of the scheme now able to analyse noisy dynamics with time-dependent generators from a sequence of snapshots. We hence provide a diagnostic tool for a wider spectrum of instances while inheriting all the favourable features from the previous protocol.

On the theoretical side, the problem of characterising time-dependent Markovian channels has been an open problem for many decades. This work gives an approach to tackle this characterisation problem rigorously.
\end{abstract}

\section{Introduction}

In the current era of noisy, intermediate-scale quantum devices we face the major challenge of error control and mitigation.
This is necessary both in order to carry out computation on near term devices which are not invalidated by noise, and to work towards the goal of building large scale fault tolerant devices.
A number of procedures have been developed to inform about noisy effects in different settings.
Markovian evolutions, that is, dynamics not retaining any memory of their history, have been investigated with particular interest.
Indeed, when characterising error models it is often taken as an underlying assumption that the error processes are Markovian~\cite{helsen2020general,Knill_2008,Magesan_2011,merkel2019randomized,Wallman_2018}.
A relevant question, therefore, is whether it is possible to establish if a quantum channel is compatible with a Markovian dynamics.
This is sometimes referred to as the \emph{quantum embedding problem} which has been shown to be NP-hard in general~\cite{CEW09,Bausch16}, but efficiently (classically) computable for any fixed Hilbert space dimension.
Ref.~\cite{Markdynamics08} illustrates a theoretical solution in terms of a semi-definite integer program checking whether there exists a branch of the matrix logarithm of the input satisfying three precise necessary and sufficient conditions to a Lindbladian operator.
While the solution in \cite{Markdynamics08} is theoretically interesting, its practical utility is limited since it is not well suited to handle inputs which are experimentally implemented unitary channels, nor to allow for imprecision in the measurement phase.
In ref.~\cite{noMarkpaper21} we proposed a novel scheme bringing the work in~\cite{Markdynamics08} into a fully functional algorithm based on convex optimisation which is able to analyse arbitrary channels in realistic noisy scenarios.
The algorithm in \cite{noMarkpaper21} takes as input a tomographic \emph{snapshot} of a quantum channel (i.e. a transfer matrix obtained via process tomography on the dynamics of interest), checks whether it is compatible with Markovian evolution, and in the affirmative case,  extracts a full description of the Lindbladian which best fits the tomographic data. The complete characterisation of the closest Lindblad operator to the generator of the real dynamics can then be used to design error correction and mitigation strategies.

A remaining drawback of the methods presented in \cite{noMarkpaper21} is that they only test for, and fit noise models to, time-\emph{independent} Markovian dynamics.
While on short time scales it is reasonable to assume that the evolutions are not varying with time,  extending the methods to treat the case of time-dependent dynamics is key to widening their applicability.
Hence, in this work we present a new algorithm, based on the techniques in~\cite{noMarkpaper21}, which now searches for a compatible time-\emph{dependent} Markovian evolution.
The algorithm uses convex optimisation techniques and leverages the continuous complete positivity property of Markovian channels to find the best fit Lindblad generators to a small number of tomographic snapshots, where we no longer require that the Lindbladian is invariant over the whole evolution.
Hence we can obtain a complete quantitative description of a time-dependent Markovian dynamics, or alternatively give solid evidence of a non-Markovian evolution when the method fails in retrieving a valid sequence of Lindbladians compatible with the tomographic input data.

The algorithm presented here inherits the favourable properties of the algorithms in ref.~\cite{noMarkpaper21}.
Namely, it does not necessitate of any a priori assumption about the noise models being investigated, requires only a small number of tomographic snapshots, and is flexible in dealing with tomographic inaccuracies.
Moreover, thanks to the convex optimisation task at the core of the scheme, it can also be used to investigate dynamics with a moderate amount of non-Markovian noise.

\section{Preliminaries}

\subsection{Notation}

We denote elementary basis vectors by $\ket{e_j} = (0,\dots,1,0,\dots,0)^T$ with 1 in the $j$-th position. The maximally entangled state is $\ket{\omega}= \sum_{j=1}^d \ket{e_j,e_j}/\sqrt{d}$ and $\omega_\perp = \1 - \dyad{\omega}{\omega}$ is the projection onto its orthogonal complement.
We write $\Flip$ to denote the flip operator interchanging the tensor product of elementary basis vectors, i.e.~$\Flip \ket{e_j,e_k} = \ket{e_k,e_j}$. We will use the Frobenius norm on matrices, defined by $\Fnorm{M} = \sqrt{\sum_{j,k} \abs{m_{jk}}^2}$, and the 1-norm, $\norm{M}_1 = \sum_{j,k} \abs{m_{jk}}$. The partial trace of the first $d$-dimensional system will be denoted by $\Tr_1 [\ \cdot\ ]$.

\subsection{Channel representations}\label{sec:channel_representations}

We will consider quantum channels of finite dimension only, i.e.~completely positive and trace preserving (CPT) linear operators acting on the space of $d \times d$  matrices. To represent a channel $\mc C$ as a $d^2\times d^2$ matrix $C$, we will adopt the elementary basis representation:
\begin{equation}
	C_{(j,k),(\ell ,m)} = \bra{e_j,e_k} C \ket{e_\ell,e_m} \coloneqq \Tr \big[ \dyad{e_k}{e_j} \mc C (\dyad{e_\ell}{e_m}) \big].
\end{equation}
The corresponding representation $\ket{v}\in\C^{d^2}$ of a $d\times d$ matrix $V$ on which the channel acts is then
$v_{j,k}=\braket{e_j,e_k}{v} \coloneqq \bra{e_j}V\ket{e_k}$.
In this representation, the action of the channel on a matrix becomes matrix-vector multiplication, and the composition of channels corresponds to the product of their respective matrix representations.

In order to formulate necessary and sufficient conditions for the generator of a Markovian evolution, we will also make use of another representation, the Choi matrix (or Choi representation), defined as
\begin{equation}
	\tau (\mc C) \coloneqq d \big(\mc C \otimes \mc I) (\dyad{\omega}) \big) .
\end{equation}
Conveniently, the two representations are directly related through the\linebreak $\Gamma$-involution~\cite{Markdynamics08}, acting on the elementary basis as
\begin{equation}
	\dyad{e_j,e_k}{e_\ell,e_m}^\Gamma \coloneqq \dyad{e_j,e_\ell}{e_k,e_m}.
\end{equation}
Explicitly, we have
\begin{equation}
	\tau = C^\Gamma \qquad \text{and} \qquad C=\tau^\Gamma .
\end{equation}
The Choi-representation is very useful to investigate the hermiticity-preserving property of a channel, that is, the property of mapping the set of hermitian matrices onto itself. Indeed we have
\begin{equation}
	\text{$\mc C$ is hermiticity-preserving $\iff$ $\tau$ is hermitian.}
\end{equation}

\subsection{Matrix logarithm}

Matrices having non-degenerate spectrum can be uniquely diagonalised.
Their matrix logarithms are given by an infinite number of branches indexed by a vector $\vec{m}=(m_1,\dots,m_{d^2}) \in \mathds{Z}^{d^2}$.
The $0$-branch of the matrix logarithm of a diagonalisable matrix $M= \sum_{c=1}^{d^2} \lambda_j \ketbra{r_j}{\ell_j}$, with $\{\lambda_1,\dots\lambda_{d^2}\}$ being the eigenvalues of $M$ and $\ell_j$, $r_j$ the respective (not necessarily orthogonal) left and right eigenvectors, is given by
\begin{equation}
	L_0\coloneqq \log(M) = \sum_{j=1}^{d^2} \log \lambda_j \ketbra{r_j}{\ell_j};
\end{equation}
the $\vec{m}$-branch is then
\begin{equation}
	L_{\vec{m}} \coloneqq L_0 + 2 \pi \i \, \sum_{c=1}^{d^2} m_j \ketbra{r_j}{\ell_j}.
\end{equation}

\subsection{Quantum Markovian dynamics}\label{prem:Markovian_channel}

A (time-dependent) \emph{quantum Markov process} is a continuous completely positive map whose generator, which we will call the \emph{Lindbladian}, must obey the (time-dependent version of) the well-known \emph{Lindblad form}~\cite{lindblad1976,Gorini76,Breuer_2009},
\begin{equation}\label{eq:Lindbladian_diagonal}
	\mathfrak{L}(t,\rho)
	\coloneqq
	\i [\rho,H(t)]
	+ \sum_\alpha \gamma_\alpha
	\left[
	J_\alpha(t) \rho J_\alpha^\dagger(t) -\frac{1}{2} \left( J_\alpha^\dagger(t) J_\alpha (t) \rho + \rho J_\alpha^\dagger (t) J_\alpha (t) \right)
	\right] .
\end{equation}
where $H(t)$ is hermitian, $\gamma (t) \geq 0$ are the \emph{decoherence rates} and $\set{J_\alpha}_\alpha$ are called \emph{jump operators}.
The first term on the rhs is the Hamiltonian part and describes the unitary evolution of the density operator, while the second term represents the dissipative part of the process.
Hence, we can write a quantum Markov process from time $t_1$ to $t_2$ as a completely positive and trace preserving channel
\begin{equation}\label{eq:mark_map}
	\Phi(t_1,t_2)
	=
	\mathbb{T} \exp \int_{t_1}^{t_2} L(t) \dt
	=
	\lim_{\dt\rightarrow 0} \prod_{j=N}^1 \e^{L_j},
\end{equation}
where $\mathbb T$ is the time-ordering operator, $ N=(t_2-t_1)/\dt$ and $L_j$ are time-independent Lindbladians (in some matrix representation), such that it is divisible at any intermediate time, that is,
\begin{equation}\label{eq:divisibility_property}
	\Phi (t_1,t_3) = \Phi (t_2,t_3) \Phi(t_1,t_2) \quad \text{for all} \ 0\leq t_1\leq t_2 \leq t_3.
\end{equation}

Whether a matrix $L_j$ constitutes a legitimate Lindbladian can be established thanks to three are necessary and sufficient conditions~\cite{Markdynamics08}. In the elementary basis representation, these are
\begin{enumerate}[label=(\roman*)]
	\item $\mathcal L_j$ is hermiticity-preserving, that is, $\mc{L}(V)^\dagger = \mc{L}(V^\dagger)$ for any matrix $V$.
	In the natural basis representation, this corresponds to $L_j \,\Flip\ket{v^\ast} =\Flip(L_j \ket{v})^\ast$ for all $\ket{v}$.
	\item $(L_j)^\Gamma$ is \emph{conditionally completely positive}~\cite{Evans77}, namely,
	$\omega_\perp \, (L_j)^\Gamma \, \omega_\perp \geq 0,$
	where $\omega_\perp = (\1 - \dyad{\omega}{\omega})$.
	\item $\bra{\omega} L_j= \bra{0}$, which corresponds to the trace-preserving property.
\end{enumerate}

We can re-formulate the above condition in terms of the Choi representation $X=L^\Gamma$, that is,
	(i) $X$ hermitian,
	(ii) $\omega_\perp \,X \, \omega_\perp \geq 0,$
	(iii) $\Tr_1[X]=0_{d,d}$.

\section{Problem and setting}

We consider a Markovian quantum channel $\Phi(t)$ with $\Phi(0)=\mathcal{I}$ whose description is completely unknown, and a time series of tomographic snapshots $M_1,M_2, \dots, M_N$ thereof (in the natural basis representation) taken at different times
$t_1,t_2, \dots, t_N=\mc T$. Our goal is to extract a consistent sequence of Lindblad operators whose evolution is the best Markovian approximation to each snapshot $M_p$.
To this aim, we iteratively construct $\Theta_p = M_p M_{p-1}^{-1}$ for $p= 2, \dots , N$ with~$\Theta_1 = M_1$.
Under the  Markovian assumption, the divisibility property in \cref{eq:divisibility_property} yields that all $\Theta$s are completely positive and trace preserving maps.
In this way we translate our problem into finding separately for every time interval between two snapshots $M_{p-1}$ and $M_p$ a time-independent Lindbladian $L_p$ whose evolution $\exp L_p$ approximates the mapping $\Theta_p$. In other words, we characterise each snapshot $M_p$ as a sequence of time-independent Markovian evolutions, i.e.,
$M_p \approx \prod_{q=p}^1 \e^{L_q}$,
which will converge in the limit of infinitely small time intervals as per \cref{eq:mark_map} (we provide a quantitative expression for the bound on the error in \cref{sec:bounds}).

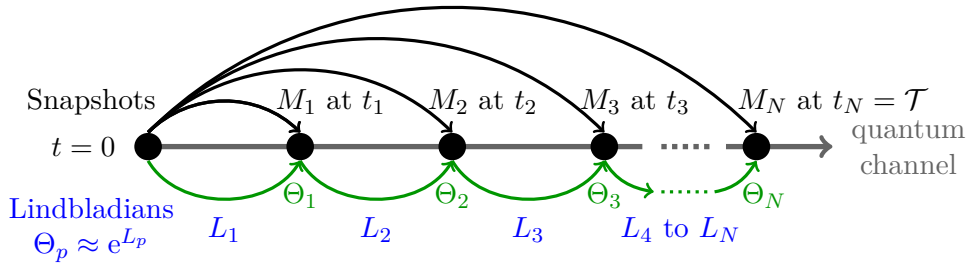
\begin{figure}[h]
	\begin{tikzpicture}

		\node at (-.75,.6) [align=center] {\color{black}{Snapshots}};
		\node at (-.75,-1.1) [align=center] {\color{blue}{Lindbladians} \\ \color{blue}{$\Theta_p \approx \e^{L_p}$}};

		\draw [line width=2pt, mygrey] (0,0) -- (6.5,0);
		\draw [dotted,line width=2pt, mygrey] (6.75,0) -- (7.4,0);
		\draw [->,line width=2pt, mygrey] (7.6,0) -- (9,0);

		\node at (10,0) [align=center] (a){\color{mygrey}{quantum} \\ \color{mygrey}{channel}};

		\node at (0,0) [fill,circle](b0) {};
		\node at (-.85,0) [align=center] {$t=0$};

		\foreach \y in {1,2,3}{
			\node at (\y*2,0) [fill,black,circle](b\y) {};
			\draw [->,very thick,black] (b0.north) to[out=45,in=135] (b\y.north);
			\node at (.4+2*\y,.6) [align=center] {\color{black}{$M_{\y}$ at $t_\y$}};
			\node at (2*\y,-.65) [align=center] {\color{mygreen}{$\Theta_{\y}$}};

			\pgfmathtruncatemacro{\z}{\y-1}

			\draw [->,very thick,mygreen] (b\z.south) to[out=300,in=240] (b\y.south);
			\node at (-1+2*\y,-1.1) [align=center] {\color{blue}{$L_{\y}$}};
		}

		\node at (8,0) [fill,black,circle](bN) {};
		\node at (9 ,.6) [align=center] {\color{black}{$M_{N}$ at $t_N=\mc T$}};
		\node at (8.1,-.65) [align=center] {\color{mygreen}{$\Theta_{N}$}};
		\draw [->,very thick,black] (b0.north) to[out=45,in=135] (bN.north);
		\draw [->,very thick,black] (b0.north) to[out=45,in=135] (b1.north);

		\node at (7,-1.1) [align=center] {\color{blue}{$L_4$ to $L_N$}};

		\draw [->,very thick,mygreen] (b3.south) to[out=300,in=170] (6.65,-.6);
		\draw [dotted,very thick,mygreen] (6.75,-.6) -- (7.4,-.6);
		\draw [->,very thick,mygreen] (7.45,-.6) to[out=10,in=240] (bN.south);

	\end{tikzpicture}
\end{figure}

\paragraph{Lipschitz continuity assumption}
We make one assumption on the time-dependent dynamics: namely that the time-dependent Lindbladian does not vary arbitrarily quickly, i.e.\ has bounded derivative. More precisely, we assume it satisfies a Lipschitz condition: $\Fnorm{L(t_2)-L(t_1)} \leq \eta (t_2 - t_1)$ with Lipschitz constant~$\eta$.
This condition corresponds physically to assuming the generator is not fluctuating arbitrarily rapidly over short time scales, which is generally justified on physical grounds.

\section{A solution based on convex optimisation}

Moving from the natural to the Choi representation, we can then formulate this strategy as a convex optimisation task obeying the three constraints illustrated in \cref{prem:Markovian_channel}.
More specifically, we want to minimise the distance $X(p,\vec m) - G(p,\vec m)^\Gamma$ between branches of the matrix logarithm of $\Theta_p$, denoted by $G(p,\vec m)$, and the variable $X(p,\vec m)$ corresponding to a Lindbladian generator in Choi representation.
The scheme then returns the set of Lindbladians $\set{L_p}_p$ generating the closest evolution to the series of snapshots $\set{\Theta_p}_p$ over a truncated set of branches.

\medskip

The algorithm implementing this approach is given schematically in \cref{alg:time_dependent}, which we explain below.

\begin{algorithm}\small
	\SetKwInOut{Input}{Input}\SetKwInOut{Output}{Output}
	\Input{integer $N$, matrices $M_1,\dots,M_N$,
		positive integer $m_{\max}$,
		positive real $\beta$ }
	\Output{set of Lindbladians  $\set{L_p}_p$ generating best-fit map to $\Theta_p = M_{p}M_{p-1}^{-1}$}

	\For{ $\vec m \in \set{m_{\max}}^{\times d^2}$}{

		\For{$p=1,\dots,N$}{
			$\Theta_p \leftarrow M_{p}M_{p-1}^{-1}$ \\
			$\set{\ell_j, r_j}_{j=1}^{d^2} \leftarrow$ set of left and right eingenvectors of $\Theta_p$ \\
			$G(p,\vec m) \leftarrow \log \Theta_p + 2\pi \i \sum_{j=1}^{d^2} m_j \, \ketbra{\ell_j}{r_j}$

			\smallskip

			Run convex optimisation programme on variable $X(p,\vec m)$:\\
			\Indp $\begin{array}{ll}
				\text{minimise} & \Fnorm{X(p,\vec m)-G(p,\vec{m})^\Gamma}   \\
				\text{subject to } &X(p,\vec m) \text{ hermitian}\\
				&\omega_\perp X(p,\vec m) \omega_\perp \geq 0 \\
				&\norm{\Tr_1[X(p,\vec m)]}_{1} = 0 \\
				&\Fnorm{X(p,\vec m)-X(p-1, \vec m)}\leq  \beta
			\end{array}$\\
			\Indm
			\textsl{Store} $X(p,\vec m)$
		}

		\textsl{Store} $\mathrm{distance}(\vec{m}) \leftarrow \sum_p \Fnorm{\Theta_p - \exp X(p,\vec m)^\Gamma}$

	}

	\textsl{set} $\set{X(p) \leftarrow X(p,\vec{m'})}_p \textsl{ for } \vec{m'} = \mathrm{argmin}\, \{\mathrm{distance}(\vec{m})\}$

	\Return $\set{L_{p}\coloneqq X(p)^\Gamma}_p$
	\caption{time-dependent Markovian map estimation}\label{alg:time_dependent}
\end{algorithm}

We remark that, conveniently, the times $t_p$ when the snapshots $M_p$ are taken is not required input.

We also note that we are limiting the search of the best-fit Lindbladians over a restricted number of branches of $\log \Theta$ up to the index~$m_{\max}$.
From previous numerical analysis on both synthesised and experimental data,  we indeed observed that a search with $m_{\max}=1$ is typically enough to obtain a good fit.
However, if needed techniques do exist to search up to the theoretical maximum efficiently:
it is known that setting $m_{\max} =  \mc O(2^{2^{\poly(d)}})$ will always suffice to find a solution, if one exists \cite{Khachiyan+Porkolab}.

A crucial caveat is the analysis of channels having degenerate spectrum in their ideal, noiseless form (this is for instance the case for any quantum unitary).
Indeed, the ill conditioning of basis vectors for multi-dimensional eigenspace with respect to perturbation results in a direct application of the convex optimisation scheme (as well as the strategy of previous work~\cite{Markdynamics08}) failing to recognise a Lindbladian generating a map being close to the measured channel.
We have solved this crucial issue with a pre-processing algorithm detecting clusters of eigenvalues stemming from a degenerate one, and adjusting the respective eigenvectors in order to restore the hermiticity-preserving structure of the input matrices.
This strategy is effective in guaranteeing an extraction of Lindblad operators close to the measured data.
An in-depth explanation of the problem, a detailed pre-processing algorithm along with a rigorous mathematical justification of this approach constitute one of the main technical contributions of ref.~\cite{noMarkpaper21}; we refer the reader to that work for further details.

Along with the conditions required of a legitimate Lindbladian operator, in the constraints of the convex optimisation task we make the assumption that the Lindbladian is moderately slowly varying, and as such include a Lipschitz continuity bound -- characterised by the parameter $\beta \equiv \beta(\eta)$ -- for the difference between two consecutive generators. (Note that the Frobenius norm is invariant with respect to the $\Gamma$ involution.)
This ensures that even though we are not taking measurements at infinitesimally short time intervals, the resulting channel $\widetilde{M} = \Pi_{p=1}^Te^{L_p}$ is close to the true channel in the case of a time-dependent Markovian evolution.
This is made rigorous by calculating error bounds in the resulting channel with respect to the number of snapshots taken (see \cref{sec:bounds}).
When the snapshots are taken at regular intervals $\mc T /N$, we have
 \begin{equation}
 	\beta(\eta) = \eta \frac{\mc T^2}{N^2} + 4 \mathfrak{R},
 \end{equation}
where $ \mathfrak{R}$ is the error for the truncation of the Magnus expansion of $\Phi(t)$ at the first order (cfr. \cref{sec:Lipschitz_bound}). 

In a more practical sense, when using the algorithm one can start with an estimate for $\beta$ and then vary this parameter over multiple runs, observing how the output changes.
Increasing the value of $\beta$ allows for a search over a larger space around the Lindbladian obtained in the previous iteration of the $p$-loop (that is, the generator of the previous time interval), which will likely give a better fit to the most suited branch of~$\log \Theta_p$.
From a physical perspective, if the distance is significantly reduced when augmenting $\beta$, this suggests that the Lindblad generator is varying more rapidly than anticipated.
Conversely, obtaining good fits to the data when setting $\beta$ with small values suggests that the process is close to time-independent dynamics.
Against intuition, relaxing the bound on the difference between two consecutive Lindbladians can instead increase the distance between $M_p$ and its Markovian estimation $\Pi_{j=1}^p L_j$, but not beyond explicitly derived bounds~\cite{CEW09}. The underlying reason for this behavior is that there is no direct correspondence between distance of two matrices and their logarithms.

\section{Error bounds}

\subsection{Estimation of time-dependent Markovian channels}\label{sec:bounds}

We want to investigate the error in the approximation of the time series snapshots of a time-dependent Markovian quantum channel with best-fit Lindbladians from our scheme, and in particular to link the degree of accuracy with the number of snapshots taken. In other words, we want to establish how fine-grained the tomographic measurements should be in order to achieve the desired level of accuracy.

To simplify calculations and presentation, we assume that all snapshots are taken at uniform time intervals $\frac{\mc T}{N}$ over a total run time $\mc T$, but we note that the results can be extended to the general case straightforwardly.
As a second assumption for $\Phi(t)$, we consider an evolution where the Lindbladian is time-dependent but with moderately varying fluctuation; as discussed, we impose a Lipschitz continuity
$\norm{L(t_2)-L(t_1)} \leq \eta (t_2 - t_1)$ for some Lipschitz constant~$\eta$.
We then have

\begin{theorem}[Snapshots approximation]
	Let $\set{M_p}_{p=1}^N$ be a time series of $N$ tomographic snapshots over a total run time $\mc T$ of a Markovian quantum channel acting on a $d$-dimensional space. Let $\eta$ be the Lipschitz constant of the generator for some norm $\norm{\ \cdot \ }$.
	Construct $\Theta_p=M_p M_{p-1}^{-1}$ and define $\wt M_p=\prod_{j=p}^1 \e^{L_j}$ with best-fit Lindbladians $\set{L_p}_p$ from \cref{alg:time_dependent}.
	Then for all $p=1,\dots, N$,
	\begin{equation}\label{eq:thm_expression_1}
		\norm{\Theta_p - \e^{L_p}} \leq \eta^2 \frac{\mc{T}^4}{N^3} + \mc O \Big(\frac{\mc T^5}{N^4}\Big)
	\end{equation}
	and
	\begin{equation}
		\norm{M_p - \wt M_p}
		\leq
		\sqrt d \Big[ \exp\Big(\sqrt d \, \eta^2 \frac{\mc{T}^4}{N^2}\Big) - 1 \Big] + \mc O \Big(\frac{\mc T^5}{N^3}\Big) .
	\end{equation}
\end{theorem}
The number of snapshots should hence scale quadratically with respect to the run time, and the difference between snapshots and reconstructed best-fit channel closes in the asymptotic limit $N\rightarrow \infty$, in line with the analysis presented in~\cite{WolfDivisibility}.

\begin{proof}
	We first investigate a bound on $\Theta_p$.
	Consider the time-averaged Lindbladian
	$L_\avg(p) = \frac{N}{\mc T} \int_{(p-1)\frac{\mc T}{N}}^{p\frac{\mc T}{N}} L(t) \dt$ and expand the time-ordered integral as a Dyson series,
	\begin{align}
		\Theta_p= &\mathbb T \exp \int_{(p-1)\frac{\mc T}{N}}^{p\frac{\mc T}{N}} L(t) \dt \\
		&=
		\1 + \int_{(p-1)\frac{\mc T}{N}}^{p\frac{\mc T}{N}} L(t) \dt
		+
		\int_{(p-1)\frac{\mc T}{N}}^{p\frac{\mc T}{N}} dt_1 \int_{(p-1)\frac{\mc T}{N}}^{t_1} dt_2 \, L(t_1) L(t_2) \\
		&+
		\int_{(p-1)\frac{\mc T}{N}}^{p\frac{\mc T}{N}} dt_1 \int_{(p-1)\frac{\mc T}{N}}^{t_1}  dt_2 \int_{(p-1)\frac{\mc T}{N}}^{t_2} \dt_3 \, L(t_1) L(t_2) L(t_3) + \dots \\
		&=
		\1 + L_\avg(p) \frac{\mc T}{N}
		+
		\sum_{k=2}^\infty
		\int_{(p-1)\frac{\mc T}{N}}^{p\frac{\mc T}{N}} dt_1 \int_{(p-1)\frac{\mc T}{N}}^{t_1} dt_2 \dots \int_{(p-1)\frac{\mc T}{N}}^{t_{k-1}} \dt_k \, L(t_1) L(t_2) \cdots L(t_k). \label{eq:Dyson_unfolded}
	\end{align}
	Let us consider the second Dyson term, and note that thanks to the Lipschitz continuity in the observed time interval  we can write
	$L(t) = L_\avg(p) + \eta \frac{\mc T}{N} Y(t)$ for some $Y(t)$ with $\norm{Y(t)} \leq 1$. Taking the norm yields
	\begin{align}
		&\norm{\int_{(p-1)\frac{\mc T}{N}}^{p\frac{\mc T}{N}} dt_1 \int_{(p-1)\frac{\mc T}{N}}^{t_1} dt_2 \, L(t_1) L(t_2)} \\
		&\leq
		\int_{(p-1)\frac{\mc T}{N}}^{p\frac{\mc T}{N}} dt_1 \int_{(p-1)\frac{\mc T}{N}}^{t_1} dt_2 \, \norm{(L_\avg(p) + \eta \frac{\mc T}{N} Y(t_1))(L_\avg(p)+ \eta \frac{\mc T}{N} Y(t_2)) } \\
		&\leq
		\frac 1 2 \norm{L_\avg(p)}^2 \frac{\mc{T}^2}{N^2} + 2 \eta \norm{L_\avg(p)} \norm{\max Y(t)} \frac 1 2 \frac{\mc{T}^3}{N^3} + \eta^2 \norm{\max Y(t)}^2 \frac 1 2 \frac {\mc{T}^4}{N^4}\\
		&\leq
		\frac 1 2 \norm{L_\avg(p)}^2 \frac{\mc{T}^2}{N^2} + 2 \eta \norm{L_\avg(p)} \frac 1 2 \frac{\mc{T}^3}{N^3} + \eta^2 \frac 1 2 \frac {\mc{T}^4}{N^4}.
	\end{align}
	In the same vein, the norm of $k$-th term (with $k\geq 2$) in the Dyson series can be bounded as
	\begin{align}
		&\norm{\int_{(p-1)\frac{\mc T}{N}}^{p\frac{\mc T}{N}} dt_1 \int_{(p-1)\frac{\mc T}{N}}^{t_1} dt_2 \dots \int_{(p-1)\frac{\mc T}{N}}^{t_{k-1}} \dt_k \, L(t_1) L(t_2) \cdots L(t_k)} \\
		&\leq
		\frac {1}{k!} \norm{L_\avg(p)}^k \frac{\mc{T}^k}{N^k}
		+ k \eta \norm{L_\avg(p)} \frac {1}{k!}\frac{\mc{T}^{k+1}}{N^{k+1}}
		+ \dots\\
		&=
		\sum_{\ell=0}^k \binom k \ell \norm{L_\avg(p)}^{k-\ell}\eta^\ell \frac {1}{k!} \frac{\mc{T}^{k+\ell}}{N^{k+\ell}} \\
		&=
		\sum_{\ell=0}^k \frac{1}{\ell!} \eta^\ell \left(\frac{\mc{T}}{N}\right)^{2\ell}
		\frac{1}{(k-\ell)!} \norm{L_\avg(p)}^{k-\ell} \frac{\mc{T}^{k-\ell}}{N^{k-\ell}}.
	\end{align}
	Now if we fix $\ell$ and sum over $k$ (that is, we sum the $\ell$-th item in the sum for the bound of each Dyson term) we get
	\begin{align}
		&\frac{1}{\ell!} \eta^\ell \left(\frac{\mc T}{N}\right)^{2\ell}
		\sum_{k\geq \ell}\frac{1}{(k-\ell)!} \norm{L_\avg(p)}^{k-\ell} \frac{\mc{T}^{k-\ell}}{N^{k-\ell}} \\
		&=
		\frac{1}{\ell!} \eta^\ell \left(\frac{\mc T}{N}\right)^{2\ell}
		\sum_{m\geq 0}\frac{1}{m!} \norm{L_\avg(p)}^{m} \frac{\mc{T}^{m}}{N^{m}} \\
		&=
		\frac{1}{\ell!} \eta^\ell \left(\frac{\mc T}{N}\right)^{2\ell} \exp \left(\norm{L_\avg(p)} \frac{\mc T}{N} \right), \label{eq:bound_for_ell}
	\end{align}
	and summing now \cref{eq:bound_for_ell} over $0\leq \ell \leq k $ gives
	\begin{equation}\label{eq:two_exponentials}
		\exp \left(\eta \frac{\mc{T}^2}{N^2}\right) \exp \left( \norm{L_\avg(p)} \frac{\mc T}{N}\right) .
	\end{equation}
	Note that in \cref{eq:two_exponentials} we have accounted for an extra term not appearing in \cref{eq:Dyson_unfolded}, that is, the item for $k=\ell=1$ that should thus be subtracted, yielding
	\begin{equation}
		\norm{\mathbb T \exp \int_{(p-1)\frac{\mc T}{N}}^{p\frac{\mc T}{N}} L(t) \dt}
		\leq
		\exp \left(\eta \frac{\mc{T}^2}{N^2}\right) \exp \left( \norm{L_\avg(p)} \frac{\mc T}{N}\right) - \eta \frac{\mc{T}^2}{N^2}.
	\end{equation}

	For the actual bound on the approximation of $\Theta_p$ we thus have
	\begin{align}
		\norm{\Theta_p - \exp L_p}
		&\leq
		\norm{\mathbb T \exp \int_{(p-1)\frac{\mc T}{N}}^{p\frac{\mc T}{N}} L(t) \dt - \exp L_p}  \\
		&\leq
		\norm{\mathbb T \exp \int_{(p-1)\frac{\mc T}{N}}^{p\frac{\mc T}{N}} L(t) \dt - \exp \Big(L_\avg(p) \frac{\mc T}{N}\Big)}\\
		&\leq
		\left(\exp \eta \frac{\mc{T}^2}{N^2}  - 1\right) \exp \Big(\norm{L_\avg(p)} \frac{\mc T}{N}\Big) - \eta \frac{\mc{T}^2}{N^2} \\
		&=
		\eta \frac{\mc{T}^2}{N^2} \left( \exp \Big(\norm{L_\avg(p)} \frac{\mc T}{N}\Big) - 1 \right)
		+ \mc O (\eta^2 \frac{\mc{T}^4}{N^4}).
	\end{align}
	Considering that by Lipschitz continuity we have $\norm{L_\avg(p)}\leq \eta \mc T$, neglecting terms $\mc O (\frac{\mc T^5}{N^4})$ we obtain the first result in the theorem in \cref{eq:thm_expression_1},
	\begin{equation}\label{eq:bound_T_p_approximation}
		\norm{\Theta_p - \exp L_p} \leq \eta^2 \frac{\mc{T}^4}{N^3} .
	\end{equation}


	\medskip

	Now we turn to the error bound in the snapshot approximation, which is clearly the largest at the latest snapshot $M_N$.
	By writing $\Theta_p =   \exp L_p  + A_p$ we then have
	\begin{equation}
		\norm{M_N - \wt M_N}
		=
		\Big\Vert \prod_{p=N}^1 (\exp L_p  + A_p) - \prod_{p=N}^1 \exp L_p  \Big\Vert .
	\end{equation}
	Now, following the argument in~\cite{WolfDivisibility}, we note that $\prod_{p=N}^1 (\exp L_p  + A_p)$ contains $\binom{N}{k}$ terms with $N-k$ exponentials~$\exp L_p$ and $k$ operators from~$\set{A_p}_p$, where the former come in at most $k+1$ separated groups each of those bound by $\sqrt d$~\cite[Theorem 1]{contractivityPT}, and the latter according to \cref{eq:bound_T_p_approximation} are bound by
	$ \max \norm{A_p} \leq \eta^2 \frac{\mc{T}^4}{N^3} + \mc O (\frac{\mc T^5}{N^4})$ .
	Hence,
	\begin{align}
		\norm{M_N - \wt M_N}
		&\leq
		\sqrt d \Big[ \Big(1+\sqrt d \, \eta^2 \frac{\mc{T}^4}{N^3} + \mc O (\frac{\mc T^5}{N^4})\Big)^N - 1 \Big] \\
		&\leq
		\sqrt d \Big[ \exp \Big( \sqrt d \, \eta^2 \frac{\mc{T}^4}{N^2} \Big) - 1 \Big] + \mc O (\frac{\mc T^5}{N^3}) .
	\end{align}

\end{proof}

\subsection{Bound on difference between consecutive extracted Lindbladians}\label{sec:Lipschitz_bound}
In this section we prove the following result for the interpretation of the parameter $\beta(\eta)$ in \cref{alg:time_dependent}.
This can be expressed in relation to the Lipschitz constant~$\eta$ of the Lindblad operator, the total run time~$\mc T$ and the total number of snapshots taken, $N$. More precisely, $\beta(\eta)$ is characterised by the RHS of \cref{eq:bound_for_beta} of the following theorem.

\begin{theorem}[Lipschitz bound]
	Consider a time-dependent Markovian process $\Phi(t)$ characterised by a Lipschitz continuous Lindblad generator $L(t)$ with Lipschitz constant $\eta$ with respect to some norm $\norm{\ \cdot \ }$.
	 Then the set of best-fit Lindbladians $\set{L_p}_{p=1}^N$ to
	 \begin{equation}
	 	\log \Phi( (p-1)\mc T/N,p \mc T/N ) \quad p=2,\dots,N
	 \end{equation}
 	satisfies
	 \begin{equation}\label{eq:bound_for_beta}
	 	\norm{L_p - L_{p-1}} \leq \eta \, \mc T^2 / N^2 + 2 \Big(\mathfrak R (p) + \mathfrak R (p-1) \Big)
	 \end{equation}
 	where $\mathfrak R(p)$ is the error for the truncation at the first order of the Magnus series over the interval $[t_{p-1},t_p]$.
\end{theorem}

\noindent In the specific case of snapshots taken at regular intervals $\mc T/N$, we thus have:
\begin{equation}
	\beta(\eta) = \eta \frac{\mc T^2}{N} + 4 \mathfrak R .
\end{equation}

\begin{proof}
	We consider the operator $K_p = \log \Phi( (p-1)\mc T/N,p \mc T/N )= L_{\avg} (p) \frac{\mc T}{N} + \mathfrak R (p)$, which is not guaranteed to be of Lindbladian form~\cite{Schnell20,Schnell21}. By triangle inequality we have
	\begin{align}
		\norm{L_p - L_{p-1}}
		&\leq
		\norm{L_p - K_p + K_p - L_{p-1} + K_{p-1} - K_{p-1} } \\
		&\leq
		\norm{L_p - K_p} + \norm{ L_{p-1} - K_{p-1} } + \norm{ K_p - K_{p-1} } \label{eq:bound_triangle}.
	\end{align}
	Since $L_p$ is to be considered the best-fit Lindbladian and observing that the first order Magnus term $L_\avg (p) \frac{\mc T}{N} = \int_{(p-1)\mc T /N}^{p \mc T /N}   L(t) \dt$ is a Lindbladian, it follows that
	\begin{equation}
		\norm{L_p - K_p} \leq \norm{L_\avg (p) \frac{\mc T}{N} - K_p} \eqqcolon \mathfrak R (p),
	\end{equation}
	and equivalently for  $\norm{ L_{p-1} - K_{p-1} }$.
	For the third term in \cref{eq:bound_triangle}, we write
	\begin{align}
		\norm{ K_p - K_{p-1} }
		&\leq
		\norm{L_\avg (p) \frac{\mc T}{N} - L_\avg (p-1) \frac{\mc T}{N} } + \mathfrak R (p) + \mathfrak R (p-1) \\
		&=
		\norm{ \int_{(p-1)\mc T /N}^{p \mc T /N} L(t)  \dt - \int_{(p-2)\mc T /N}^{(p-1) \mc T /N} L(t)  \dt} + \mathfrak R (p) + \mathfrak R (p-1) \\
		&=
		\norm{ \int_{(p-1)\mc T /N}^{p \mc T /N} \Big( L(t) - L(t-\mc T/N) \Big)  \dt } + \mathfrak R (p) + \mathfrak R (p-1) \\
		&\leq
		\eta \, T /N \int_{(p-1)\mc T /N}^{p \mc T /N} \dt + \mathfrak R (p) + \mathfrak R (p-1) \\
		&=
		\eta \, T^2 /N^2 + \mathfrak R (p) + \mathfrak R (p-1) .
	\end{align}
\end{proof}
\section{Conclusions}

In these notes we have extended our previous method~\cite{noMarkpaper21} to fit noise models to tomography data compatible with time-dependent Lindbladian evolutions.
Departing from the time-independent assumption is a significant improvement to analyse quantum maps in a real-world scenario, and it widens the applicability of \cite{noMarkpaper21} to assess quantum dynamics affected by more complicated noise patterns.
The cost of this approach is the higher number of snapshots, which is required to scale quadratically in the total run time in order to guarantee a given estimation accuracy, while in \cite{noMarkpaper21} a single snapshot sufficed.
However, necessitating additional snapshots is unavoidable when investigating potentially time-dependent dynamics.

We view the new time-dependent scheme presented here, and the previous time-independent scheme, as complementary to each other in practice, rather then just viewing the former as a refinement of the latter.
Especially for small time scales, we expect both approaches to successfully fit the observed data, where the time-independent algorithm returning a single Lindbladian may be more convenient for experimentalists to design error mitigation techniques. Conversely, for longer run times one would expect the time-dependent version to successfully identify a set of Lindbladians fitting the data for any continuous Markovian process affected by noise varying over time, while the time-independent algorithm certifies stable evolutions driven by a constant Lindblad generator.

\section*{Acknowledgements}
E.O. is supported by the Bavarian state government with funds from the Hightech Agenda Bayern Plus as part of the Munich Quantum Valley, and by the UK Hub in Quantum Computing and Simulation, part of the UK National Quantum Technologies Programme with funding from UKRI EPSRC (grant EP/T001062/1).
T.K. is supported by the Spanish Ministry of Science and Innovation through the ``Severo Ochoa Program for Centres of Excellence in R$\&$D'' (CEX2019-00904-S) and PID2020-113523GB-I0 and started the work while being supported by the EPSRC through the Centre for Doctoral Training in Delivering Quantum Technologies (grant EP/L015242/1).
This work was supported by Google Research Award ``Assessing non-Markovian noise in NISQ
devices''.
T.S.C. was supported by the Royal Society.
This work has been supported in part by the EPSRC Prosperity Partnership in Quantum Software for Simulation and Modelling (grant EP/S005021/1), and by the UK Hub in Quantum Computing and Simulation, part of the UK National Quantum Technologies Programme with funding from UKRI EPSRC (grant EP/T001062/1).


\end{document}